\documentclass{emulateapj}
\usepackage{apjfonts}
\newcommand{\lsun}{\rm L$_\odot$}

\shorttitle{Organic Dust
 in the HR 4796A Disk} 
\shortauthors{Debes et al.}

\begin{document}
\title{Complex Organic Materials in the Circumstellar Disk of HR 4796A}
\author{John H. Debes\altaffilmark{1},Alycia J. Weinberger\altaffilmark{1},Glenn Schneider\altaffilmark{2}}

\altaffiltext{1}{Department of Terrestrial Magnetism, Carnegie Institution of 
Washington, Washington, DC 20015}
\altaffiltext{2}{Steward Observatory, The University of Arizona, Tucson, AZ 85721}
\vfill

\begin{abstract}

We combine HST/NICMOS imaging photometry of the HR 4796A disk at previously unobserved wavelengths between 1.71-2.22\micron\ with reprocessed archival observations to produce
a measure of the dust's scattering efficiency as a function of wavelength.  
The spectrum of the dust, synthesized from the seven photometric measures,
is characterized
by a steep red slope increasing from 0.5~\micron\ to 1.6~\micron\ followed by a flattening
of the spectrum at wavelengths $>$ 1.6~\micron.  We fit the spectrum
with a model population of dust grains made of tholins, materials comprised of
complex organic materials seen throughout the outer parts of our Solar System.
The presence of organic material around a star that may be in the later stages
of giant planet formation implies that the basic building blocks for life may be common in planetary systems.
\end{abstract}


\section{Introduction}
HR 4796A is a young (8 Myr) star
with a circumstellar debris disk that has been imaged at several 
wavelengths \citep{rayjay98,koerner98,schneider99,schneider06}.  
Dust in the disk scatters starlight at visible to near-IR wavelengths
and emits thermally in the infrared.  Both scattered and thermal emission
are spatially coincident originating in a $<$ 17~AU wide annulus \citep{rayjay98,schneider99,koerner98,schneider06} 70~AU from the central star.
Estimates of the
mass of material present in the disk range from
$10^{26}-10^{27}$ g \citep{jura93,jura95}.  The constituent grains 
assumed to be responsible for the spatially resolved scattered light and thermal
emission have radii on
the order of a few \micron\, although 
unresolved emission at infrared and sub-mm wavelengths require larger grains a few tens of \micron\ in size
to be present
\citep{augereau99,li,wahhaj04,kessler05}.
The smaller grains will 
escape from the HR 4796A system on short timescales implying a resevoir
of colliding bodies that exists to fuel the dust production. 
  The characteristics of the HR 4796A 
disk make it a younger, dustier analogue to the Kuiper Belt within the Solar System.  Studying HR 4796A provides a probe of planet
formation processes around a star with a mass and effective temperature
significantly different than our own.

The ring structure of the disk was inferred by thermal emission
well fit by a single blackbody temperature \citep{jura93,jura98}.  
It was recognized that 
the disk could contain dust that was not simply made of silicates, but
available observations could not usefully constrain the true 
composition of the dust grains \citep{augereau99,wahhaj04}.  Mid-IR Keck
LWS spectra from
8-13\micron\ similarly showed a lack of observable silicate lines or other 
identifying features \citep{kessler05}.
Near-IR scattered light images of the circumstellar dust \citep{schneider99} suggested a red spectral slope at these wavelengths, hinting at a composition
similar to that seen around Solar System bodies in the Kuiper Belt. 

\section{Observations}
\label{sec:obs}
 In order to measure the reflectivity of the dust, we observed
HR 4796A with the Hubble Space Telescope's (HST) Near-Infrared
and Multi-Object Spectrometer (NICMOS) \citep{thompson98} 
using the coronagraph in Camera 2.  We
used four medium bandwidth filters ($\Delta\lambda/\lambda\simeq10\%$), F171M, F180M, F204M, and F222M, with $\lambda_c$= 1.71, 1.80, 2.04, and 2.22 \micron. 
 
All new observations were taken on 09 May 2005
 for HR~4796A and HR~4748, a PSF reference.  
The observations include direct
images of both stars with short exposures for stellar photometry, as 
well as longer exposures for coronagraphic high contrast imaging. 
The instrumentally calibrated and reduced images
were created from the raw NICMOS {\it multiaccum} exposures following
the processing methodolgy  described by \S3 of \citet{schneider05} and 
references therein.  Additionally, we took archival 
raw images in the STIS 50CCD ($\lambda$=0.585), NICMOS F110W ($\lambda$=1.1\micron), 
and F160W filters ($\lambda$=1.6\micron) and re-reduced the images.

For each of the medium bandwidth filters we directly measured HR 4796A and HR 4748's
flux densities with their uncertainties
to obtain wavelength dependent brightness ratios for PSF subtraction. Due to the (detector saturating) brightness, unocculted stellar
images were not available in the broader passbands. Thus, we estimated the 50CCD, F110W, and
F160W photospheric flux densities with synthetic photometry using
 the CALCPHOT task of the SYNPHOT\footnote[1]{http://www.stsci.edu/resources/software\_hardware/stsdas} package in conjunction with Kurucz models tabulated in the
SYNPHOT library.  We used a solar metallicity, $\log{g}$=4.5 model of a
9250~K star, using a luminosity of 21.1 \lsun\ from
a best fit match to HR~4796A
based on its parallax from the Hipparcos satellite, its Tycho-2 $B$ and $V$
photometry, as well as its 2MASS $J$, $H$, and $K_s$ photometry.  For HR 4748
we used data from the same sources to derive a 121 \lsun, $\log{g}$=4.5 model with
T$_{eff}$=10250 K.  We compared the predicted F171M, F180M, F204M, and F222M photometry with the SYNPHOT values--they agreed to within a few percent of our 
measured flux densities.  We fit the long wavelength emission from
the disk to a single temperature blackbody and found T$_{eff}$=78$\pm$1~K and $L_{IR}/L_\star$=4.8$\pm0.1 \times10^{-3}$, a 
useful surrogate for the disk optical depth.  Since $L_{IR}/L_\star\ll 1$,
it is reasonable to assume that the disk is optically thin.  However, if the ring had a small enough scale height, the disk's radial optical depth would approach unity.  We discuss the implications of a radially optically thick disk on our proposed model in \S\ref{sec:spectrum}.

In order to determine the best subtraction we minimized a chi-square metric on a 
region of the diffraction spikes for the target star.  We assume that good
subtraction of the diffraction spikes corresponds to the best subtraction
of the PSF within the region of interest \citep{cnc03}.  We searched within 1-$\sigma$ of
our scaling uncertainty and iteratively searched to within 0.1 pixels of the
best offsets to get the optimal subtraction between target star and PSF.

To quantify the systematic effects on the photometry, we repeated the subtractions varying the PSF scalings and offsets by $\pm$1 $\sigma$ from the minimum chi-square solution found above. Using an elliptical photometric aperture matched to the size and shape of the disk, we found the standard deviation in the disk flux densities from this suite of subtractions and used this as a systematic uncertainty per pixel that was then propagated into the disk 
photometry.

We observed HR 4796A in each medium band filter at two celestial orientations differing by
26$^\circ$.  This is essentially an azimuthal dither that allows true
objects that rotate with the field of view to be distinguished from rotationally invariant
instrumental artifacts.  The two
differentially-oriented PSF-subtracted images in each filter band were geometrically rectified and corrected for the linear optical distortion (X$_{scale}$= 75.95 mas pixel$^{-1}$, Y$_{scale}$=75.42 mas pixel$^{-1}$) at the Camera 2 coronagraphic focus.  The geometrically corrected image pairs were rotated about the location of the occulted target (as determined through the target acquisition process), then co-aligned to higher precision using the image centroids of a nearby field star common in both images, and, finally,  combined by
averaging.  

\begin{figure}[t]
\begin{center}
\scalebox{0.8}{\plotone{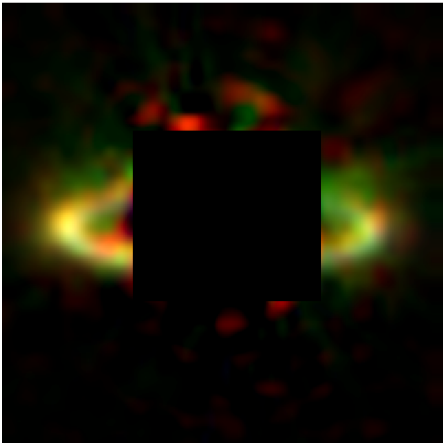}}
\end{center}
\caption{\label{fig:colim}False color image of the HR 4796A disk with three of the
passbands imaged.  Blue corresponds to the 50CCD image, green 
is the F110W image, and red is the F171M image.  A slight color difference between the two lobes of the disk is seen.
The central region
where PSF subtraction errors dominate is masked out.}
\end{figure}

To determine the brightness of the disk from the PSF subtracted images, we chose an elliptical
aperture with a major-to-minor axis ratio equal to the measured inclination
of the disk of 14.1$^\circ$ from edge-on \citep{schneider06}.  We used an inner major axis
boundary of 0$\farcs$8, and an outer major axis boundary 
 of 1$\farcs$5.  At $r<$0\farcs8, systematic errors dominate the image, which restricts useful photometry of the disk to within $\pm$20$^\circ$ of each ansa.  Beyond 1\farcs5, 
disk flux is below 5-10\% of the peak.
 We used an elliptical annulus
between 1$\farcs$9 and 2$\farcs$5 with the same azimuthal coverage to determine the background level of the image.
   Table \ref{tab:res} gives 
the total measured flux densities of the disk for the four filters.

There were three differences with our re-reduction and subtraction of the PSF star for the Cycle 7 F110W and F160W data.  We directly calculated pixel centers of the stars behind the coronagraphic hole using the acquisition images and slews reported in
the data file headers.  A flight software problem caused the relevant 
header keywords of the acquisition image centroid to be incorrectly populated.
The geometric distortion of
Camera 2 is time dependent, 
so we used X$_{scale}$=75.87 mas pixel$^{-1}$ and Y$_{scale}$=75.19 mas pixel$^{-1}$,
appropriate for these observations.

\begin{deluxetable*}{cccccc}
\tablecolumns{6}
\tablewidth{0pc}
\tablecaption{\label{tab:res} Properties of the HR 4796A disk}
\tablehead{
\colhead{Filter} & \colhead{Scaling\tablenotemark{a}} & \colhead{Flux Density} & \colhead{Total Disk Flux Density} & \colhead{NE:SW ratio} & \colhead{$g$} \\
 & & \colhead{ (mJy)} & \colhead{ (mJy)} & & \\
}
\startdata
STIS 50CCD & 0.705 & 5.5$\pm$0.2 & 9.4$\pm$0.8 & 0.74$\pm$0.07 & 0.16$\pm$0.06 \\
F110W & 0.78$\pm$0.03 & 5.2$\pm$0.3 & 8.1$\pm$0.6 & 0.9$\pm$0.08 & 0.06$\pm$0.07 \\
F160W & -\tablenotemark{b} & 3.7$\pm$0.5 & 5.7$\pm$0.9 & 1.0$\pm$0.1 & 0.0$\pm$0.06 \\
F171M & 0.819$\pm$0.007 & 2.8$\pm$0.1 & 4.8$\pm$0.5 & 0.62$\pm$0.08 & 0.3$\pm$0.1 \\
F180M & 0.79$\pm$0.01 & 2.9$\pm$0.1 & 4.7$\pm$0.3 & 0.71$\pm$0.07 & 0.2$\pm$0.1 \\
F204M & 0.79$\pm$0.02 & 2.5$\pm$0.1 & 4.2$\pm$0.2 & 1.02$\pm$0.09 & 0.2$\pm$0.1 \\
F222M & 0.79$\pm$0.03 & 2.1$\pm$0.1 & 3.5$\pm$0.2 & 1.0$\pm$0.1 & 0.1$\pm$0.1 \\
\enddata
\tablenotetext{a}{Ratio of HR 4796A to HR 4748}
\tablenotetext{b}{No contemporaneous PSF reference, used five different references with different scalings}
\end{deluxetable*}

We measured disk asymmetries from our new and reprossessed images, as has been done
previously for 50CCD and F110W \citep{schneider06,augereau99}.  
The disk appears to be asymmetric about the major axis, in a manner consistent with forward scattering dust grains.  We quantify this asymmetry in the context
of the Henyey-Greenstein $g$ parameter.  Additionally, the northeastern (NE) side of the disk is
brighter than  the southwestern (SW) side.
  We present both sets of measurements in Table \ref{tab:res}.  There is a 
marginal trend to a larger $g$-value that peaks at 1.71\micron.  The asymmetry between the two sides of the disk is particularly strong at 1.71 and 1.8\micron

Figure \ref{fig:colim} shows a three-color image of HR 4796A using information from the STIS,
F110W, and F171M filters to highlight visually the spectrum of the disk and any spatial variations in the spectrum.

To determine the assumed forward scattering asymmetry of the ring,
we measured the total flux in the four readily visible quadrants (between 5
and 15$^\circ$
from the major axis) of the disk between 0\farcs8 and 1\farcs5,
and took a ratio between the front (brighter) and back (dimmer) parts of the NE and SW ansae of the
 disk.
We calculated the same ratios using a Henyey-Greenstein phase
function \citep{henyey41} for the disk inclination to determine
a best fit value for the asymmetry $g$ at each wavelength.  
To measure the brightness asymmetry between the two lobes, we
took the total flux in the NE and SW lobes and measured their ratios
($\pm20^\circ$ from the major axis) between 0\farcs8 and 1\farcs5, using the
same background annulus.

It is useful to determine how the observed surface brightness in an optically
thin disk relates to dust properties.  The average surface brightness ($F_{\nu,disk}/A_{disk}$) of an optically thin disk is the product of the 
total scattering cross section 
of the dust integrated over the 
volume of the disk multiplied by the flux density of the star (F$_{\nu,\star}$) and the phase
function of the dust $\Phi(g,\theta)$ \citep{augereau06,schneider06b}.   Since
our observations were close to the ansae, we assumed that the phase function
of the dust could be approximated by the Henyey-Greenstein phase function at 
a scattering angle of $\sim90^\circ$ and assuming the median $g$ value for all 
of the wavelengths.  The measured asymmetries may be used to calculate a geometrical factor relating the total disk flux to the observed flux close to the ansae.  The geometrical factor
was calculated by using an azimuthally averaged radial cut of both lobes of the
disk ($\pm20^\circ$ from the major axis) and by generating a model disk from that measurement.  Additionally, we multiplied the model disk by a 
sine function to approximate the observed NE to SW asymmetry.  We measured the ratio of the total flux in the model disk to that of the apertures we used to determine the geometrical factor at each wavelength.

\begin{figure}[b]
\plotone{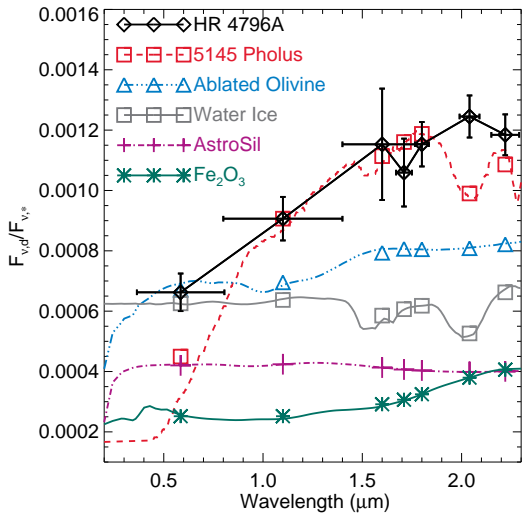}
\caption{\label{fig:spec}$F_{\nu,disk}/F_{\nu,\star}$ for HR 4796A.  Horizontal bars indicate the FWHM of each bandpass.  The red curve and symbols represent
5145 Pholus.  The other curves represent grain models for candidate materials with an $n^{-3.5}$ size distribution with $a_{min}$=3\micron\ and $a_{max}$=1000\micron.  Symbols represent the models integrated over each bandpass. Each curve is normalized to the F110W data of HR 4796A and offset for clarity.}   
\end{figure}

\section{HR 4796A's Disk Spectrum}
\label{sec:spectrum}
A combination of the surface brightness and $L_{IR}/L_\star$ in conjunction 
with Mie scattering models of a given dust distribution and composition can
self-consistently provide a measure of the scattering
efficiency ($\left<Q_{sca}\right>$) of the HR 4796A disk:

\begin{equation}
\label{eq:final}
\left<Q_{sca}\right>=4 \frac{F_{\nu,disk}}{<a^2>A_{disk}F_{\nu,\star}}\frac{(1+g^2)^{1.5}}
{(1-g^2)}\left[\frac{\int{\left<\sigma_{abs}\right>F_{\nu,\star}d\nu}}{\int{F_{\nu,\star}d\nu}}\right]\frac{L_\star}{L_{IR}}
\end{equation}

where $A_{disk}$ is the area of the disk, $g$ is the asymmetry parameter for the dust, $\pi\left<a^2\right>$ is the effective grain cross section given a power law dust distribution with a minimum and maximum grain size, and $\left<\sigma_{abs}\right>$
is the effective absorption cross section per grain of the dust.

 Figure \ref{fig:spec} shows $F_{\nu,disk}/F_{\nu,\star}$
spanning from the visible to the near-IR.  The red slope of the disk is 
reminiscent of organic materials on the surface of Centaur 
5145 Pholus but not of other expected components of planetesimals or common
red materials such 
as water ice, astronomical silicates, hematite\footnote[2]{Optical constants from http://www.astro.uni-jena.de/Laboratory/OCDB/data/oxide/hem\_interp.dat}, or olivine
reddened by UV ablation \citep{cruikshank98,cuzzi02,astrosil,pollack94,brunetto07}, also plotted in Figure \ref{fig:spec}.  
The scattering properties
of organic materials formed in the outer Solar System
are not well known, but optical constants exist
for tholins, complex organic polymers that
 reproduce the spectral characteristics of
organic materials in the atmosphere of Titan and which have been synthesized on Earth
\citep{khare83}. Tholins have been used to successfully
reproduce the red spectra of icy Solar System bodies \citep[i.e.][]{cruikshank98}.  Similar bodies may be colliding to 
create the dust around HR 4796A. 

\begin{figure}[t]
\epsscale{.7}
\plotone{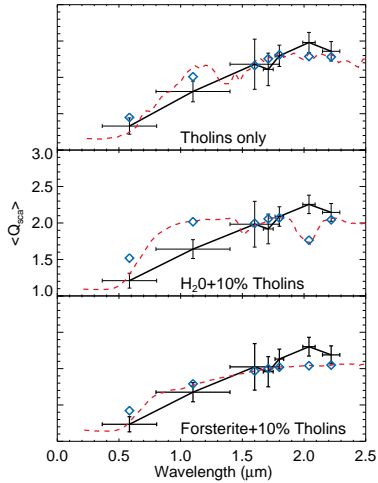}
\caption{\label{fig:mod}Comparison of three models of the dust disk
around HR 4796A--(top) tholins with a$_{min}$=1.4\micron, a$_{max}$=980\micron,
(middle) water ice with tholin inclusions that comprise 10\% of the volume with a$_{min}$=12 \micron,a$_{max}$=100\micron, (bottom) forsterite with tholin inclusions that comprise 10\% of the volume with a$_{min}$=12\micron,a$_{max}$=110\micron.}
\end{figure}

The top panel in Figure \ref{fig:mod} shows a comparison of the disk spectrum with a best fit grain model of Titan tholins only with two free parameters: a$_{min}$=1.4\micron\, a$_{max}$=980\micron\ and a fixed power
law index of $\kappa$=-3.5 \citep{dohnanyi69}, with a reduced chi-squared value of the fit of 1.2.
The chi-squared contours imply that minimum grain sizes of between 1\micron\ and  5\micron\ are possible within the 98\% confidence contour.  The spectrum is weakly dependent on
the maximum grain size and so is not a meaningful constraint on the maximum size of the dust.  This is true for all the models shown.
 
Given that the HR~4796A disk's thermal emission can be modeled with
a 78~K blackbody, it is conceivable that methane, water, or some
other ice may be present in the disk \citep[but see][]{artymowicz}.  The spectral shape we see may be
 due to the presence of tholin inclusions in a larger ice or silicate matrix,
since pure ice and silicate models do not possess the red slope
seen in the data.
Using effective optical constants for water ice with tholin inclusions \citep{bohren} and 
forsterite with tholin inclusions we compare the resulting 
$\left<Q_{sca}\right>$ with our data in Figure 4.  The reduced chi-squared value
for these two mixtures is 6.8 and 2.0, respectively.  These values make an
ice+tholin mixture unlikely, but do not rule out the presence of tholin
and silicate mixtures. 

We note that the general red spectral 
shape of the disk is very similar to 5145 Pholus, one of the redder
objects in the Solar System, which is believed to be comprised of water ice with tholin inclusions mixed with olivine and methanol ice
\citep{cruikshank98,cuzzi02}.  
Additionally, we confirm 
that the minimum grain size in our models for 
the scattered light is comparable in size to the 
minimum grain size
predicted by a detailed fit of the long wavelength SED \citep{wahhaj04}.
Tholins provide a useful proxy for the dust present in HR 4796A, but may not
be the exact substance observed, given the large number of different compounds
that make up tholins.

If, as discussed in \S\ref{sec:obs} the disk is marginally
optically thick in the radial direction, the overall choice of composition 
would remain unchanged, though the details of grain size and distribution may
be different.  In a multiple scattering case, the surface brightness of the disk would depend on the extinction efficiency of the dust and $g$ in addition
to $\left<Q_{sca}\right>$.  The extinction efficiency and $g$ don't appreciably vary as a function of wavelength betwee 0.5-2\micron\ for tholin grains. Therefore, even in the case of multiple scattering, the 
scattering spectrum would depend solely on $\left<Q_{sca}\right>$.  Given that the total mass
of dust is uncertain by an order of magnitude and the scale height of the
disk is unknown, the optically thin assumption is justified.

The calculations we performed also make strong predictions on the forward 
scattering behavior of the disk grains.  The scattering phase
asymmetry parameters predicted by our 
models for tholin dust are much higher ($g\sim$0.8) than actually observed (See Table \ref{tab:res}).  
For spherical grains, $g$ is strongly dependent on grain
 size relative to observed wavelength, and more weakly dependent on 
composition.
The measured asymmetries are much smaller than predicted by our grain model.
It is possible that the shape and porosity of the grains make the assumption of spherical symmetry for Mie scattering incorrect and thus markedly changes
the phase function of the dust.  Porous spherical grains or fractal 
aggregates tend to have
higher $g$ values \citep{wolff98,bertini07}, however, so it is not clear
what mechanism is responsible for the lower values of $g$ observed.  Other disks seem to show low asymmetry
\citep{schneider06b}.

\section{Conclusions}
We have measured a visible to near-IR photometric reflectance spectrum
of the dust ring present around HR 4796A.  The scattered light from the ring
is suggestive of a grain population dominated by 1.4\micron-radius grains of complex organic material similar
in optical properties to the material observed in the atmosphere of Titan and 
the surfaces of icy bodies in the Solar System.  The model grains fit the scattered light data but do not fit the observed azimuthal asymmetries in the observed brightness of the dust ring, suggesting that
the grains are not spherical.

The library of optical constants for material is not wholly complete, but tholins
provide the best fit of the current selection of materials as illustrated
in Figure \ref{fig:spec}.  Many minerals and ices can be summarily rejected because they possess relatively neutral scattering efficiencies from 0.5-1.5\micron.  UV ablated olivine has been used to explain particularly red 
asteroids due to space weathering \citep{brunetto07} and hematite is found
on the surface of Mars, but these materials cannot reproduce the observed slope we see in the regime of Mie scattering.  Thus, the presence of organic material is the most plausible explanation for the observations.

Longer wavelength scattered light observations will further constrain the grain
models we have used, 
particularly around 3.8-4\micron\ where a large absorption feature is
seen for different grain sizes of tholins.  This would help to directly 
confirm whether Titan tholins are an adequate proxy for the material
in orbit around HR 4796A.  Additionally, measuring the optical properties of 
organic materials in meteorites and from samples of the Stardust mission will
provide further tests of our model grains.

\acknowledgements
This research is based on observations with the NASA/ESA Hubble Space Telescope which is operated by the AURA, under NASA contract NAS 5-26555. These observations are associated with program \#'s 7233, 8624, and 10167.  Support for programs \#8624 and \#10167 was provided by NASA through a grant from STScI.  AJW also acknowledges support from the NASA Astrobiology Institute.  J.D wishes to thank Jeff Cuzzi for insightful conversations on Mie scattering and Alexandra Surcel for a careful reading of the manuscript.

\end{document}